\documentclass{elsarticle}
\usepackage{geometry}
\usepackage{array,multirow}
\usepackage{graphicx}
\usepackage{amsmath}
\usepackage{amsfonts}
\usepackage{amssymb}
\usepackage{subcaption}
\usepackage{hyperref}
\usepackage{siunitx}
\usepackage{booktabs}
\usepackage{xurl}
\usepackage{lineno}
\usepackage[color=cyan!20,textsize=scriptsize]{todonotes}
\newif\ifnotes
\notestrue

\newcommand{\pp}{p\text{-value}}

\journal{Physica A}

\begin{document}
\begin{frontmatter}
\title{Emergence of Randomness in Temporally Aggregated Financial Tick Sequences}

\author[1]{Silvia Onofri}
\ead{silvia.onofri@sns.it}
\author[2]{Andrey Shternshis\corref{cor1}}
\ead{andrey.shternshis@it.uu.se}
\author[1]{Stefano Marmi}
\ead{stefano.marmi@sns.it}
\cortext[cor1]{Corresponding author}

\affiliation[1]{organization={Scuola Normale Superiore},
addressline={Piazza dei Cavalieri 7},
postcode={56126},
city={Pisa},
country={Italy}}

\affiliation[2]{organization={Uppsala University},
addressline={Regementsvägen 10},
postcode={75105},
city={Uppsala},
country={Sweden}}

\begin{abstract}
Market efficiency implies that price changes are unpredictable, so that an efficient market behaves like a random number generator. At the level of individual transactions, however, prices are dominated by microstructural effects and are far from random. We study how randomness emerges when tick-by-tick prices are aggregated in transaction time. The signs of aggregated returns are encoded as binary strings, whose randomness is assessed with the NIST Statistical Test Suite, the Alphabit and Rabbit batteries of TestU01 and two entropy-based tests, after a sanity check that selects the tests suited to the available string lengths. A solvable benchmark explains the whitening: for stationary Gaussian models of tick returns with positive long-run variance, the entropy deficit of the sign sequence vanishes as the inverse square of the aggregation level, and in a finite sample the residual memory remains detectable up to a level growing as the cube root of the number of transactions. Data on nine US assets confirm that randomness increases with aggregation, at a pace that depends on trading activity, and that stocks with a large relative tick size are most predictable at intermediate aggregation levels. The procedure yields an online, model-free source of pseudorandom bits.
\end{abstract}

\end{frontmatter}


\section{Introduction}
\label{intro}

A central concept in finance, the Efficient Market Hypothesis (EMH) \citep{Fama}, posits that asset prices fully reflect all available information, ruling out systematic risk-adjusted excess returns based on historical data. The intuition behind market efficiency has long motivated comparing price fluctuations to pure noise; as \cite{hasanhodzic2011computational} state, ``the more efficient the market, the more random the sequence of price changes generated by it, and the most efficient market of all is one in which price changes are completely random and unpredictable.'' Motivated by this perspective, we treat an efficient market as a random number generator (RNG).

This analogy has been taken literally. Doyle and Chen \cite{doyle2013patterns} applied the Overlapping Serial Test of RNG batteries to the sequences of up and down moves of stock-market indices and found recurring, market-specific patterns. Moews \cite{moews2024random} treated binary daily returns of Nasdaq-listed stocks as the output of a hardware RNG and found that their efficiency varies across years, in particular during financial crises. In the opposite direction, Clark and Hengartner \cite{10.5555/1924892.1924895} proposed stock prices as a source of public randomness (a randomness beacon), an idea recently revisited in the presence of an active attacker \cite{Landis25}, and random numbers have been extracted from cryptocurrency prices \cite{chiba2024random}.

At the shortest time scales, however, prices are not random. Tick-by-tick data reflect the bid--ask bounce, the discreteness of prices and the activity of trading algorithms \cite{roll84, biais1995empirical, hasbrouck2007}, and they display strong dependence and anomalous, non-diffusive behaviour \cite{Cont, Bouchaud03}. Econometricians treat these effects as microstructure noise and look for the sampling frequency that best separates it from the efficient price \cite{ait2005often, bandi2008microstructure}, while mathematical finance models prices at coarser scales as semimartingales, the standard stochastic framework underlying no-arbitrage pricing theory \cite{delbaen2006general, ait2014high}. This paper studies the transition between the two descriptions through the lens of coarse-graining \cite{mantegna1995scaling}: we ask how randomness emerges when transactions are aggregated, how fast, and at which scales financial time series become indistinguishable from random sequences.

\textbf{Relevant literature.} The degree of randomness of financial prices is not constant. It varies across time \citep{Lo04, Mensi} and across market classes \citep{Risso1}, and it depends on the temporal resolution. Intraday prices display well-documented stylised facts \cite{Cont}; prices sampled every minute already deviate from randomness \citep{Calcagnile}, and tick-by-tick data reveal structured patterns reflecting correlated trading activity \cite{Lillo04, Bouchaud09}. Randomness tests have been used as diagnostics of market efficiency and of deterministic trading behaviour \cite{hsieh1991chaos, scheinkman1989nonlinear, Zhang18, noakes2016testing}, and information-theoretic measures \cite{Shannon, Kullback} have been employed to quantify the randomness and complexity of financial time series \cite{Dionisio, Alvarez21, Carbone22}. In a previous study \cite{shternshis2025price}, entropy-based tests showed that the predictability of ultra-high-frequency prices decreases as tick data are aggregated over larger numbers of transactions. What has been missing is a systematic assessment with standard RNG test batteries, together with a theory of why, and how fast, aggregation produces randomness.

\textbf{Contributions.} First, we treat binarised price changes as the output of an RNG and apply the NIST Statistical Test Suite \cite{Bassham_Rukhin_Soto_Nechvatal_Smid_Barker_Leigh_Levenson_Vangel_Banks_etal._2010}, the Alphabit and Rabbit batteries of TestU01 \cite{testu01,testu01_userguide} and two entropy-based tests to tick data at aggregation levels from 1 to 100 transactions. A preliminary sanity check on true and pseudo-random sequences selects the tests that are reliable at the available string lengths. Second, we derive a solvable benchmark for the whitening of sign sequences under aggregation. For stationary Gaussian models of tick returns with positive long-run variance, the entropy deficit vanishes as $\kappa^2/\ell^2$, where $\ell$ is the aggregation level and $\kappa$ a single model-dependent number; in a finite sample the residual memory remains detectable up to an aggregation level growing as the cube root of the number of transactions. The theory also identifies when the whitening is non-monotonic or fails. Third, the data confirm the predicted whitening, show that the aggregation needed to pass the tests grows with trading activity, and reveal that stocks with a large relative tick size are most predictable at intermediate aggregation levels, as the theory predicts for prices dominated by the bid-ask bounce. The lag-one autocorrelation of the signs confirms this mechanism and provides estimates of $\kappa$ which, together with the length of the records, account for the ordering of the assets; for the most active assets, however, the memory of the signs decays more slowly than the benchmark predicts. Finally, the procedure provides an online, model-free way of generating pseudorandom bits from market data.

\textbf{Outline.} Section~\ref{Theoretical Foundations} develops the theory: section \ref{subsec:binary_mapping} briefly introduces aggregation and binarisation. Section \ref{subsec:ar1} explains whitening using autocorrelations of returns and applies the theory to three examples: AR, MA, and ARMA models. Section \ref{subsec:finite} discusses the effect of finite sequence size and gives a relationship between sequence size and the whitening effect. Possible deviations in the model preventing whitening are provided in Section \ref{subsec:departures}. Section \ref{subsec:regimes} gives an overview of three regimes: microscopic, pseudorandom, and macroscopic. The paper focuses on identifying aggregation that leads to the pseudorandom regime. Section~\ref{Data description} describes the data used in the study. We introduce the taxonomy of randomness tests in Section \ref{Hypothesis testing and taxonomy} Section, entropy-based tests in  Section \ref{Entropy-based tests}, and randomness batteries in Section \ref{nist_tests}. We present the results for the sanity check in \ref{sec:sanitycheck} and the results on tick data in Section \ref{subsec:experiments}. Section~\ref{Random numbers from financial time series} discusses random number generation, and Section~\ref{Conclusion and future work} concludes the paper.

\section{Theory: whitening of sign sequences by temporal aggregation}

\label{Theoretical Foundations}

\subsection{Aggregation and binarisation}
\label{subsec:binary_mapping}

Let $P_i$ be the price of the $i$-th transaction of an asset and $X_i=\ln P_i-\ln P_{i-1}$ the tick-by-tick log-return. At aggregation level $\ell\geq1$ the log-return over $\ell$ consecutive transactions is the block sum
\begin{equation}
S^{(\ell)}_n=\sum_{i=(n-1)\ell+1}^{n\ell}X_i=\ln\frac{P_{n\ell}}{P_{(n-1)\ell}}\,,\qquad n=1,2,\dots,
\label{eq:block_sum}
\end{equation}
and its sign defines the binary sequence
\begin{equation}
B^{(\ell)}_n=
\begin{cases}
1 & \text{if } S^{(\ell)}_n>0\,,\\
0 & \text{if } S^{(\ell)}_n<0\,,
\end{cases}
\qquad
\xi^{(\ell)}_n=2B^{(\ell)}_n-1\in\{-1,+1\}\,.
\label{eq:binary_map}
\end{equation}
This is the encoding used in our tests (Section~\ref{sec:dataset_and_tests}), where each of the $\ell$ strings obtained at level $\ell$ starts from a different transaction. Blocks with $S^{(\ell)}_n=0$ arise only because prices are discrete; they are discarded in the data and have probability zero in the continuous models below. Since the blocks do not overlap, consecutive symbols share no tick return: with overlapping windows sampled at every transaction, adjacent symbols would share $\ell-1$ returns and the sequence would become more, not less, predictable as $\ell$ grows.

\subsection{Short-memory models: a universal $\ell^{-2}$ law}
\label{subsec:ar1}

Let $(X_i)$ be a stationary process with autocovariance $\gamma(k)$, with $\sum_k |k|\,|\gamma(k)|<\infty$ and long-run variance $\sigma^2_{\rm LR}=\sum_{k\in\mathbb Z}\gamma(k)>0$. Then the properties of autocovariance yield
\begin{equation}
\operatorname{Var}S^{(\ell)}_n=\sum_{|k|<\ell}(\ell-|k|)\,\gamma(k)\,,\qquad
\operatorname{Cov}\big(S^{(\ell)}_n,S^{(\ell)}_{n+m}\big)=\sum_{|k-m\ell|<\ell}\big(\ell-|k-m\ell|\big)\,\gamma(k)\,,\quad m\geq1\,.
\label{eq:block_cov}
\end{equation}
If $\gamma$ decays exponentially, $|\gamma(k)|\leq C r^{|k|}$ with $r<1$, as for every stationary ARMA($p,q$) process, the lag-one correlation of the block sums is, for $\ell\gg\max\big(|\kappa|,1/\ln(1/r)\big)$,
\begin{equation}
\rho_\ell\equiv\operatorname{corr}\big(S^{(\ell)}_n,S^{(\ell)}_{n+1}\big)=\frac{\kappa}{\ell-2\kappa}+O\big(r^{\ell}/\ell\big)\,,\qquad
\kappa=\frac{1}{\sigma^2_{\rm LR}}\sum_{k\geq1}k\,\gamma(k)\,,
\label{eq:rho_kappa}
\end{equation}
while the correlations at lags $m\geq2$ involve only $\gamma(k)$ with $k>(m-1)\ell$ and are $O(r^{(m-1)\ell}/\ell)$. Aggregation thus has two distinct effects: correlations beyond the first lag are suppressed exponentially, and the lag-one correlation, generated by the tick returns straddling the boundary between adjacent blocks, decays only as $1/\ell$. The parameter $\kappa$, a signed memory time measured in transactions, quantifies this boundary effect.

For centred Gaussian processes $\mathbb P(B^{(\ell)}_n=1)=1/2$, and the pairwise sign correlations follow from the arcsine law for clipped Gaussian processes \cite{vanvleck1966spectrum},
\begin{equation}
q_m\equiv\mathbb E\big[\xi^{(\ell)}_n\xi^{(\ell)}_{n+m}\big]=\frac{2}{\pi}\arcsin r_\ell(m)\,,
\label{eq:arcsine}
\end{equation}
where $r_\ell(m)$ is the lag-$m$ correlation of the block sums, and the full law of the signs is given by Gaussian orthant probabilities. Randomness is measured by the Shannon entropy rate \cite{Shannon,cover2006elements}
\begin{equation}
h(\ell)=\lim_{L\to\infty}\big(\mathcal H_L-\mathcal H_{L-1}\big)\,,
\label{eq:entropy_rate}
\end{equation}
where $\mathcal H_L$ is the entropy, in bits, of $L$ consecutive symbols; $h(\ell)=1$ if and only if the symbols are independent fair coin tosses. We expand the probability function of $\mathbf{\xi} = (\xi_1,\xi_2,\dots,\xi_L)$ around uniform probability $P_0(\mathbf{\xi})=2^{-L}$

\begin{equation*}
    P(\mathbf{\xi})=P_0(\mathbf{\xi})\left(1+\sum_{1\le i< j \le L}q_{j-i}\xi_i\xi_j+O(q^2)\right)
\end{equation*}

and use the Taylor expansion of block entropy $H_L$  

\begin{equation*}
\label{eq_HL}
H_L = L -     \frac{1}{2\ln{2}}\sum_{\mathbf{\xi}}P_0(\mathbf{\xi})\left(\sum_{1\le i< j \le L}q_{j-i}\xi_i\xi_j\right)^2+O(q^3).
\end{equation*}

The non-diagonal terms in the sum squared are vanished, and it gives $h(\ell)=1-(2\ln2)^{-1}\sum_{m\geq1}q_m^2+O(q^3)$. Combining it with Eqs.~\eqref{eq:rho_kappa}, \eqref{eq:arcsine}, and $\arcsin\rho_\ell\simeq\rho_\ell=r_\ell(1)\simeq\frac{\kappa}{\ell}$,
\begin{equation}
1-h(\ell)\simeq\frac{C_\kappa}{\ell^{2}}\,,\qquad C_\kappa=\frac{2\kappa^2}{\pi^2\ln2}\,.
\label{eq:universal}
\end{equation}
Such short-memory Gaussian models with $\kappa\neq0$ therefore whiten at the same rate, $\ell^{-2}$, and asymptotically the model enters only through $\kappa$. Three examples illustrate the general result.

\begin{figure}[tbp]
    \centering
    \includegraphics[width=\linewidth]{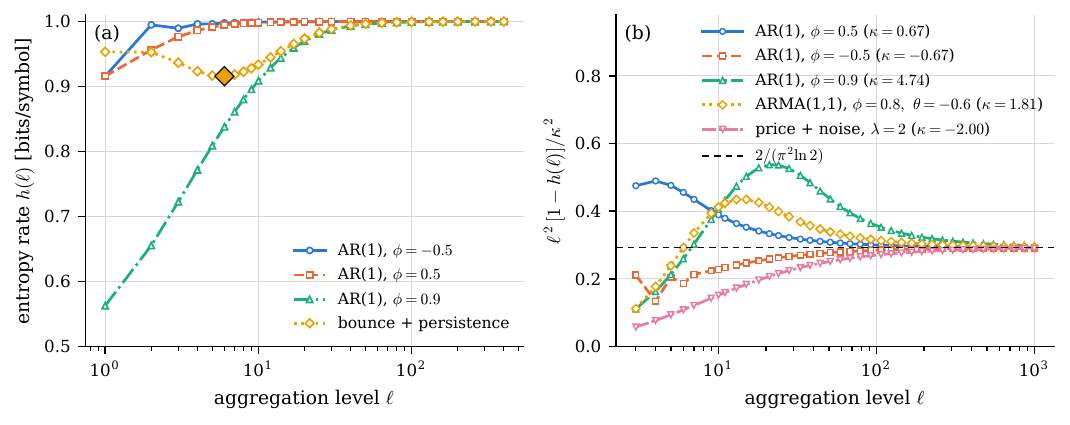}
    \caption{Entropy rate of the sign of tick returns aggregated over $\ell$ transactions, computed from exact block probabilities. (a) $h(\ell)$ for AR(1) returns with $\phi=-0.5,0.5,0.9$ and for a superposition $X_i=Y_i+u_i-u_{i-1}$ of a persistent AR(1) component ($\phi_Y=0.9$, unit innovations) and bid--ask bounce noise ($\sigma_u^2=3\gamma_Y(1)$), whose tick-level lag-one autocorrelation is $-0.28$; for the latter, predictability is maximal at $\ell=6$ (filled marker). At $\ell=1$ the AR(1) curves for $\phi=\pm0.5$ coincide by symmetry. (b) Rescaled deficit $\ell^2[1-h(\ell)]/\kappa^2$ for AR(1), ARMA(1,1) and efficient-price-plus-noise models, which converges to $2/(\pi^2\ln2)$ (dashed line), Eq.~\eqref{eq:universal}. Entropy rates are approximated by the upper bounds $h_L=\mathcal H_L-\mathcal H_{L-1}$ with $L=6$--$8$; the truncation error is below $10^{-2}$ bits (largest for the superposition model at $\ell=1$).}
    \label{fig:theory}
\end{figure}

\begin{itemize}
\item \emph{AR(1).} For $X_i=\phi X_{i-1}+\varepsilon_i$, $\kappa=\phi/(1-\phi^2)$ and, exactly \cite{amemiya1972effect,tiao1972asymptotic},
\begin{equation}
\rho_\ell=\frac{\phi\,(1-\phi^\ell)^2}{\ell(1-\phi^2)-2\phi(1-\phi^\ell)}\,.
\label{eq:rho_ell}
\end{equation}
For $\phi<0$, a crude model of the bid--ask bounce, the factor $(1-\phi^\ell)^2$ is smaller for even than for odd $\ell$: blocks with an even number of transactions largely cancel the bounce. Predictability is therefore non-monotonic at small aggregation levels; for $\phi=-0.5$ it is larger at $\ell=3$ than at $\ell=2$ (Fig.~\ref{fig:theory}a).
\item \emph{Efficient price plus noise.} For $X_i=e_i+u_i-u_{i-1}$, with independent efficient-price increments $e_i$ and noise terms $u_i$ of variances $\sigma_e^2$ and $\sigma_u^2$ \cite{roll84, ait2005often}, Eq.~\eqref{eq:block_cov} gives exactly $\rho_\ell=-\lambda/(\ell+2\lambda)$, with $\lambda=\sigma_u^2/\sigma_e^2$ the noise-to-signal ratio, so that $\kappa=-\lambda$. Aggregation whitens the signs once $\ell\gg\lambda$; the same ratio controls the optimal sampling frequency of realised variance \cite{ait2005often, bandi2008microstructure}.
\item \emph{ARMA($p,q$).} Any stationary ARMA model, such as the ARMA description of microstructure noise used in \cite{Calcagnile}, obeys Eq.~\eqref{eq:universal} with $\kappa$ computed from its autocovariance.
\end{itemize}
For these models the aggregated sign sequence is a hidden Markov chain, whose hidden state is the last tick-level state of each block, and block entropies can be computed from exact block probabilities by a forward recursion on a grid; for the superposition model of Section~\ref{subsec:departures} we integrate the Gaussian orthant probabilities numerically \cite{genz1992numerical}. Figure~\ref{fig:theory}b shows the resulting data collapse: for AR(1), ARMA(1,1) and noise models with very different $\kappa$, the rescaled deficit $\ell^2[1-h(\ell)]/\kappa^2$ converges to the universal constant $2/(\pi^2\ln2)$. Since $\ell^2\rho_\ell^2/\kappa^2\simeq(1-2\kappa/\ell)^{-2}$, the approach is from above for $\kappa>0$ and from below for $\kappa<0$.

\subsection{Finite samples}
\label{subsec:finite}

For $\kappa\neq0$ the deficit is positive at every finite $\ell$, and whether a test detects it depends on the length of the string. A record of $N$ transactions yields strings of $n_\ell=\lfloor N/\ell\rfloor$ symbols (fewer in the data, where zero returns are discarded; $N$ should then be read as an effective length). For tests of serial dependence among the signs the relevant quantity is the non-centrality parameter
\begin{equation}
\Lambda_\ell=n_\ell\sum_{m\geq1}q_m^2\simeq2\ln2\;n_\ell\,\big[1-h(\ell)\big]\simeq\frac{4\kappa^2}{\pi^2}\,\frac{N}{\ell^{3}}\,.
\label{eq:noncentrality}
\end{equation}
For instance, for $\ell\gg1$, $\Lambda_\ell$ is the expected excess of $n_\ell\hat q_1^{\,2}$ over its null value 1, where $\hat q_1$ is the empirical lag-one autocorrelation of the signs; the entropy statistics of Section~\ref{Entropy-based tests} also have non-centralities proportional to $n_\ell[1-h(\ell)]$. Detection at level $\alpha$ requires $\Lambda_\ell$ to exceed a threshold $\Lambda^\ast$ fixed by $\alpha$, the desired power and the degrees of freedom of the test, so the residual memory becomes invisible beyond
\begin{equation}
\ell_{\rm det}\simeq\left(\frac{4\kappa^2N}{\pi^2\Lambda^\ast}\right)^{1/3}\propto N^{1/3}\,.
\label{eq:ell_det}
\end{equation}
For a month of data with $N=10^6$ transactions, $\Lambda^\ast\approx6.6$ (one degree of freedom, $\alpha=0.01$, $50\%$ power) and $|\kappa|=2/3$ (AR(1) with $|\phi|=0.5$), $\ell_{\rm det}\approx30$. The aggregation level at which a string ``becomes random'' is therefore not a property of the asset alone: at equal microstructural memory it grows with the length of the record, and hence with trading activity.

\subsection{Departures from the benchmark}
\label{subsec:departures}

\textit{Non-monotonic predictability.} Consider a persistent component $Y_i$ (for instance driven by order flow) observed through bid--ask bounce, $X_i=Y_i+u_i-u_{i-1}$. At the tick level the bounce, with its negative lag-one correlation, partly masks the persistence. Under aggregation the bounce telescopes to $u_{n\ell}-u_{(n-1)\ell}$, whose variance stays fixed while that of the persistent component grows with $\ell$, so that the persistent component dominates at intermediate $\ell$, before the $\ell^{-2}$ law takes over beyond its correlation time. Predictability is then maximal at an intermediate aggregation level, $\ell=6$ in the example of Fig.~\ref{fig:theory}a. The effect requires a strong bounce component and is therefore expected in stocks whose transaction prices are dominated by the bid--ask spread, i.e.\ with a large relative tick size \cite{Dayri2015large}.

\textit{Long memory.} If $\gamma(k)\propto k^{2H-2}$ with $H>1/2$, $\sum_kk\gamma(k)$ diverges and aggregation does not whiten: for fractional Gaussian noise the block sums are again fractional Gaussian noise, the adjacent-block correlation is $\rho_\ell=2^{2H-1}-1$ for every $\ell$, and the deficit does not depend on $\ell$. Then $\Lambda_\ell\propto N/\ell$, and $\ell_{\rm det}\propto N$ instead of $N^{1/3}$. Price returns, unlike order signs, are close to uncorrelated beyond a few transactions \cite{Lillo04, Bouchaud09}, so the short-memory benchmark should apply; the scaling of the detection level with $N$ provides an empirical test.

\textit{Drift.} A mean return $\mu$ biases the signs: $\mathbb P(B^{(\ell)}_n=1)=\Phi(\delta_\ell)$, with $\Phi$ the standard normal distribution function and $\delta_\ell=\ell\mu/\sqrt{\operatorname{Var}S^{(\ell)}_n}\simeq(\mu/\sigma_{\rm LR})\sqrt\ell$. The resulting deficit, $\delta_\ell^2/(\pi\ln2)$, grows linearly in $\ell$, so $h(\ell)$ reaches a maximum near $\ell^\ast\simeq\big(4\kappa^2\sigma^2_{\rm LR}/(\pi\mu^2)\big)^{1/3}$ and then decreases. This is the discrete counterpart of the semimartingale picture, in which $\mathbb P(\Delta p>0)\simeq\Phi\big((\mu/\sigma)\sqrt{\Delta t}\big)$ over a horizon $\Delta t$ \cite{ait2020tick}: drift and diffusion scale as $\Delta t$ and $\sqrt{\Delta t}$. For a stock with daily ratio $\mathrm{SR}_d$ between mean return and volatility and $n_d$ transactions per day, $\mu/\sigma_{\rm LR}\approx\mathrm{SR}_d/\sqrt{n_d}$, so the drift becomes relevant ($\delta_\ell\approx1$) only after about $1/\mathrm{SR}_d^{\,2}$ trading days, i.e.\ years for $\mathrm{SR}_d$ of a few hundredths. In a finite sample, the expected monobit $z$-score of a string is $(2\Phi(\delta_\ell)-1)\sqrt{n_\ell}\simeq\sqrt{2/\pi}\,(\mu/\sigma_{\rm LR})\sqrt N\,(1+\kappa/\ell)$: to leading order a drift-induced imbalance is set by the trend over the whole record and is neither created nor removed by aggregation, while for $\kappa<0$ (bounce) it grows with $\ell$ towards its asymptotic value. Measuring signs relative to the median instead of zero removes it.

\textit{Asymmetric increments.} For non-Gaussian increments the imbalance reflects the median rather than the mean of the block returns, so that an asymmetry between the sizes of up and down moves produces it even without a net trend. For independent increments with skewness $\gamma_\varepsilon$, the Edgeworth expansion gives $\mathbb E\,\xi^{(\ell)}_n\simeq-\gamma_\varepsilon/(3\sqrt{2\pi\ell})$, a deficit $\gamma_\varepsilon^2/(36\pi\ln2\,\ell)$ that decays only as $1/\ell$ and, if present, eventually dominates the $C_\kappa\ell^{-2}$ term; the corresponding monobit $z$-score decreases as $1/\ell$, and its detection level grows as $N^{1/2}$. The dependence of the imbalance on $\ell$ thus distinguishes a drift (constant $z$-score) from an asymmetry of the increments (decreasing $z$-score).

\subsection{Three regimes}
\label{subsec:regimes}

Table~\ref{tab:regimes} summarises the resulting picture. The goal of the empirical analysis is to identify, asset by asset, the pseudorandom regime, in which a market acts as an efficient random number generator.

\begin{table}[tbp]
\centering
\small
\caption{Regimes of the aggregated sign sequence predicted by the short-memory benchmark ($\kappa$: boundary parameter of Eq.~\eqref{eq:rho_kappa}; $\ell_{\rm det}$: detection level of Eq.~\eqref{eq:ell_det}; $n_d$, $\mathrm{SR}_d$: daily number of transactions and ratio between mean return and volatility). The first boundary depends on the length of the record; the trend regime requires aggregation horizons, and hence records, of the order of $1/\mathrm{SR}_d^{\,2}$ trading days.}
\label{tab:regimes}
\begin{tabular}{p{0.2\linewidth} p{0.27\linewidth} p{0.43\linewidth}}
\toprule
Regime & Aggregation level & Signature \\
\midrule
Microscopic & $\ell\lesssim\ell_{\rm det}$ & serial dependence of the signs detected by the tests, strong for $\ell\lesssim|\kappa|$; possible parity effects or maximum of predictability at $\ell>1$ \\
Pseudorandom & $\ell_{\rm det}\lesssim\ell\ll n_d/\mathrm{SR}_d^{\,2}$ & deficit $C_\kappa\ell^{-2}$ below the resolution of the tests; fair-coin behaviour \\
Macroscopic trend & $\ell\gtrsim n_d/\mathrm{SR}_d^{\,2}$ & sign bias $\Phi(\delta_\ell)\neq1/2$ and collapse of the entropy rate \\
\bottomrule
\end{tabular}
\end{table}

\section{Data}
\label{Data description}

\subsection{Dataset}

We analyse limit order book data from LOBSTER (\url{www.lobsterdata.com}) for 80 trading days between 1 August and 21 November 2022, from 9:30 to 16:00. Table~\ref{Table: UHF data} lists the nine assets: eight stocks from different sectors and the SPY ETF, which tracks the S\&P 500 index. They differ in price, volatility, number of transactions (mostly less than one second apart) and daily volume (0.4 to 12 million shares). From the LOBSTER message files, which record time (with nanosecond precision), event type, order identifier, size, price and side, we retain the executions of visible and hidden limit orders \cite{Gould13}.

The last two columns of Table~\ref{Table: UHF data} characterise the tick size. For CCL, SNAP, F and INTC, the one-cent tick exceeds 3 basis points of the price and more than 80\% of consecutive transactions occur at the same price; we refer to them as large-tick stocks \cite{Calcagnile, Dayri2015large}. For AAPL, MSFT, LLY, TSLA and SPY the tick is below 1 basis point of the price.

\begin{table}[tbp]
\centering
\footnotesize
\setlength{\tabcolsep}{3.5pt}
\caption{Assets and characteristics of prices, ordered by relative tick size. Mean price, its standard deviation, trading volume, number of transactions and average time between transactions (in seconds) are computed for each day and averaged over the 80 days. The relative tick is the tick size (\$0.01) divided by the mean price; the last column is the fraction of consecutive transactions at the same price.}
\label{Table: UHF data}
\begin{tabular}{llrrrrrrr}
\toprule
Asset & Ticker & \multicolumn{1}{c}{Mean} & \multicolumn{1}{c}{Price} & \multicolumn{1}{c}{Daily} & \multicolumn{1}{c}{Daily} & \multicolumn{1}{c}{Time between} & \multicolumn{1}{c}{Relative} & \multicolumn{1}{c}{Zero} \\
 & & \multicolumn{1}{c}{price} & \multicolumn{1}{c}{s.d.} & \multicolumn{1}{c}{volume} & \multicolumn{1}{c}{transactions} & \multicolumn{1}{c}{transactions} & \multicolumn{1}{c}{tick (bp)} & \multicolumn{1}{c}{returns} \\
\midrule
Carnival & CCL & 9.24 & 0.12 & 5,874,376 & 15,372 & 1.518 & 10.8 & 84\% \\
Snap & SNAP & 10.67 & 0.14 & 4,967,779 & 18,521 & 1.358 & 9.4 & 82\% \\
Ford Motor & F & 13.93 & 0.10 & 4,468,175 & 12,954 & 1.815 & 7.2 & 83\% \\
Intel & INTC & 30.15 & 0.20 & 7,055,642 & 38,255 & 0.595 & 3.3 & 85\% \\
Apple & AAPL & 153.47 & 0.93 & 12,184,032 & 136,136 & 0.165 & 0.65 & 74\% \\
Microsoft & MSFT & 251.78 & 1.37 & 4,529,093 & 84,342 & 0.269 & 0.40 & 63\% \\
Eli Lilly & LLY & 327.33 & 1.73 & 370,050 & 11,404 & 2.086 & 0.31 & 51\% \\
Tesla & TSLA & 388.02 & 3.81 & 8,686,354 & 178,704 & 0.127 & 0.26 & 60\% \\
SPDR S\&P 500 ETF & SPY & 390.52 & 1.56 & 9,136,137 & 95,181 & 0.246 & 0.26 & 57\% \\
\bottomrule
\end{tabular}
\end{table}

\subsection{Encoding}
\label{sec:dataset_and_tests}

Prices are converted into binary strings by comparing transactions $\ell$ steps apart. For each aggregation level $\ell=1,\dots,100$ and each starting point $j=1,\dots,\ell$, the string $\overline{b}_j$ is updated, for $i=1,2,\dots$, as
\begin{equation}
r=\frac{P_{j+i\ell}}{P_{j+(i-1)\ell}}\begin{cases}
            <1 &\text{ then } \overline{b}_j\rightarrow \overline{b}_j0\,,\\
            =1 &\text{ then }\overline{b}_j\rightarrow\overline{b}_j\,,\\
            >1 &\text{ then } \overline{b}_j\rightarrow \overline{b}_j1\,,
        \end{cases}
\label{Eq:encoding}
\end{equation}
which is Eq.~\eqref{eq:binary_map} with zero returns discarded. Using all $\ell$ starting points exploits the full information even at high aggregation levels, giving $\sum_{\ell=1}^{100}\ell=5050$ strings per asset and period. The strings of the trading days of a month are concatenated, and the tests are run on monthly strings. At $\ell=1$ their length ranges from about $3\times10^4$ bits (F, November) to $1.8\times10^6$ bits (TSLA, August) (Supplementary Tables S2--S3), and at $\ell=100$ the aggregation horizon ranges from about 15~s (TSLA) to about 4~min (F, LLY) (Supplementary Tables S4--S7). Since only past prices are used, the encoding can be computed online as new transactions arrive.

\section{Randomness tests}
\label{methods}

\subsection{Hypothesis testing and taxonomy}
\label{Hypothesis testing and taxonomy}

A randomness test takes a binary string and decides between the null hypothesis $H_0$ that the string is random and the alternative that it is not, by comparing a statistic with its distribution under $H_0$. We use the significance level $\alpha=0.01$: $H_0$ is rejected if $\pp<\alpha$ for one-sided tests, and if $\pp<\alpha/2$ or $\pp>1-\alpha/2$ for two-sided ones. The tests come from the NIST Statistical Test Suite (STS) and from the Alphabit and Rabbit batteries of TestU01, complemented by two entropy-based tests (Section~\ref{Entropy-based tests}). Table~\ref{tab:taxonomytests} groups them by the property they probe: frequency tests (balance of zeros and ones), pattern tests (repeated patterns and correlations between bits), entropy and complexity tests (compressibility and predictability), spectral tests (periodic structures) and random-walk tests (excursions of the cumulative sums).

\begin{table}[tbp]
\centering
\small{\begin{tabular}{{|p{0.12\linewidth} p{0.24\linewidth} p{0.27\linewidth} p{0.25\linewidth}|}}
\hline
\textbf{Category} & \textbf{NIST STS} & \textbf{Rabbit} & \textbf{Alphabit} \\
\hline
Frequency tests &
Frequency (Monobit),\newline
Frequency Test within a Block &
MultinomialBits\-Overlapping,\newline
HammingWeight &
MultinomialBits\-Overlapping (x4)\\
\hline
Pattern tests &
Runs Test,\newline
Longest Run of Ones in a Block,\newline
Non-overlapping Template Matching,\newline
Overlapping Template Matching,\newline
Serial Test &
ClosePairsBitMatch(x2),\newline
LongestHeadRun,\newline
PeriodsInStrings,\newline
Hamming\-Correlation(x3),\newline
Hamming\-Independence(x3),\newline
AutoCorrelation(x2),\newline
Runs Test &
Hamming\-Correlation,\newline
Hamming\-Independence (x2)\\
\hline
Entropy and complexity tests &
Binary Matrix Rank Test,\newline
Maurer's Universal Statistical Test,\newline
Linear Complexity Test,\newline
Approximate Entropy Test &
AppearanceSpacings,\newline
LinearComp,\newline
LempelZiv,\newline
MatrixRank(x3) &
-- \\
\hline
Spectral tests &
Discrete Fourier Transform &
Fourier1,\newline
Fourier3 &
-- \\
\hline
Random walks tests &
Cumulative Sums Test,\newline
Random Excursions Test,\newline
Random Excursions Variant Test &
RandomWalk1(x3)&
RandomWalk1(x2)\\
\hline
\end{tabular}}
\caption{Taxonomy of the tests of NIST STS and of the Alphabit and Rabbit batteries of TestU01.}
\label{tab:taxonomytests}
\end{table}

\subsection{Entropy-based tests}
\label{Entropy-based tests}

We use the two entropy-based tests of \cite{shternshis2025price}. Let $x_1,\dots,x_N$ be the binary string and $k=[0.5\log_2N]$ the block length \cite{Shields}. The \emph{ShannonEntropy test} divides the string into $N_b=\lfloor N/k\rfloor$ non-overlapping blocks, estimates their entropy from the empirical frequencies $\hat f_j$ of the $2^k$ possible blocks,
\begin{equation}
\hat{H}=-\sum_{j}\frac{\hat{f}_j}{N_b}\ln{\frac{\hat{f}_j}{N_b}}\,,
\label{naive}
\end{equation}
and uses that $Y_1=2N_b(k\ln2-\hat H)$ is asymptotically $\chi^2(2^k-1)$ distributed under $H_0$ \cite{Zubkov74}. The \emph{KL test} relaxes equiprobability and tests only the independence of each symbol from the preceding $k-1$ ones. With $f_{ij}$ the number of occurrences, among the $N_o=N-k+1$ overlapping blocks, of the $(k-1)$-block $a_i$ followed by the symbol $a_j$, the statistic
\begin{equation}
Y_2=2\sum_{ij}f_{ij}\ln{\frac{N_of_{ij}}{f_{i\cdot}f_{\cdot j}}}\,,\qquad f_{i\cdot}=\sum_jf_{ij}\,,\quad f_{\cdot j}=\sum_if_{ij}\,,
\end{equation}
is asymptotically $\chi^2(2^{k-1}-1)$ distributed under $H_0$ \cite{shternshis2025price}.

\subsection{NIST STS and TestU01}
\label{nist_tests}

The NIST STS \cite{Bassham_Rukhin_Soto_Nechvatal_Smid_Barker_Leigh_Levenson_Vangel_Banks_etal._2010} was designed to assess random sequences for cryptographic applications; we use version 2.1.2\footnote{In \textit{src/assess.c} we changed the type of the variable \textit{expCount} from integer to float. 
}. Our strings are long enough for 9 of its 15 tests. Each test is applied to a number of sub-sequences: following the documentation, we run the Discrete Fourier Transform and Non-overlapping Template Matching tests on sub-sequences of 1000 bits and the other tests on sub-sequences of 128 bits, with the recommended values of the remaining parameters. Short sub-sequences keep their number large, but they make the \pp{}s of several tests discrete, which can bias the second-level test of uniformity of the \pp{}s when the number of sub-sequences is large; this problem is addressed by the sanity check of Section~\ref{sec:sanitycheck}.

TestU01 \cite{testu01,testu01_userguide}, version 1.2.3, is organised in batteries; we use Alphabit (9 tests: four MultinomialBitsOverlapping tests, Hamming independence and correlation tests, and two random-walk tests) and Rabbit (26 tests, including ClosePairsBitMatch, AppearanceSpacings, LinearComp, LempelZiv, Fourier1 and Fourier3, LongestHeadRun, PeriodsInStrings, HammingWeight, AutoCorrelation, Run and MatrixRank), with standard parameters. Since the standard output only reports the failed tests, we modified the functions \textit{WritepVal} and \textit{WriteReport} in \textit{testu01/bbattery.c} to collect all \pp{}s.

\section{Results}
\label{sec:experiments}

\subsection{Sanity check}
\label{sec:sanitycheck}

We first run the tests on random sequences, to exclude tests with a high false-positive rate at our string lengths and parameters. We use three generators with different sources of randomness: the Quantis QRNG USB quantum generator \cite{quantisusb}, the Linux \textit{/dev/urandom} generator, which draws on environmental noise, and the sequence of values $\mu(n)=\pm1$ of the M\"obius function on square-free integers. The latter is deterministic but conjectured to behave like a random sequence (M\"obius randomness \cite{sarnak2011three}); its randomness was examined with similar tests in \cite{Mussardo_LeClair_2021}. Details are given in Supplementary Section S1. We generate strings of $5\times10^4$, $10^5$, $5\times10^5$ and $10^6$ bits and aggregate them as the financial data; since the output of an RNG is already binary, aggregation reduces to sub-sampling, $\{b_{j+i\ell}\}_{i\geq1}$ for $j=1,\dots,\ell$ and $\ell=1,\dots,100$. A test is excluded at a given length if more than 2\% of the strings of all three generators fail it at $\alpha=0.01$.

The ShannonEntropy and KL tests pass at all lengths, and all tests of NIST STS and Alphabit pass at $5\times10^4$ bits. At larger lengths several NIST STS tests are excluded (Frequency from $10^5$ bits; BlockFrequency, CumulativeSums, FFT and NonOverlappingTemplate from $5\times10^5$; Runs and LongestRun at $10^6$), together with the shortest random walks of Alphabit and Rabbit. In Rabbit, MultinomialBitsOverlapping is excluded at all lengths and a few other tests at some lengths (Supplementary Table S1). The financial strings are tested with the selection corresponding to their approximate length: $5\times10^4$ bits for CCL, F and SNAP, $10^5$ for INTC and LLY, $5\times10^5$ for AAPL and MSFT, and $10^6$ for SPY and TSLA.

\subsection{Randomness of aggregated tick data}
\label{subsec:experiments}

Results are shown as boxplots of $-\log_{10}p_j$ over the $\ell$ strings $j=1,\dots,\ell$ at each aggregation level, with the threshold $-\log_{10}\alpha=2$ marked by a dashed line; the strings at a given level are shifted sub-samplings of the same price path and are therefore not independent. The complete set of plots is available in our GitHub repository\footnote{\url{https://github.com/silviaonofri/Emergence-of-Randomness-in-Temporally-Aggregated-Financial-Tick-Sequences}} and a selection in the Supplementary Material. In line with Section~\ref{Theoretical Foundations}, the results can be organised around four observations:
(i) randomness increases with aggregation for all assets;
(ii) the aggregation level needed to pass the tests grows with trading activity, through the length of the strings and the memory of the signs;
(iii) large-tick stocks are most predictable at intermediate aggregation levels in some tests;
(iv) frequency imbalances, due to an asymmetry between up and down moves, make the strings of several assets fail balance-sensitive tests, and for INTC they become more significant with aggregation.

\subsubsection{Whitening and the role of the record length}
\label{subsubsec:whitening}

\begin{figure}[tbp]
    \centering
    \includegraphics[width=0.86\linewidth]{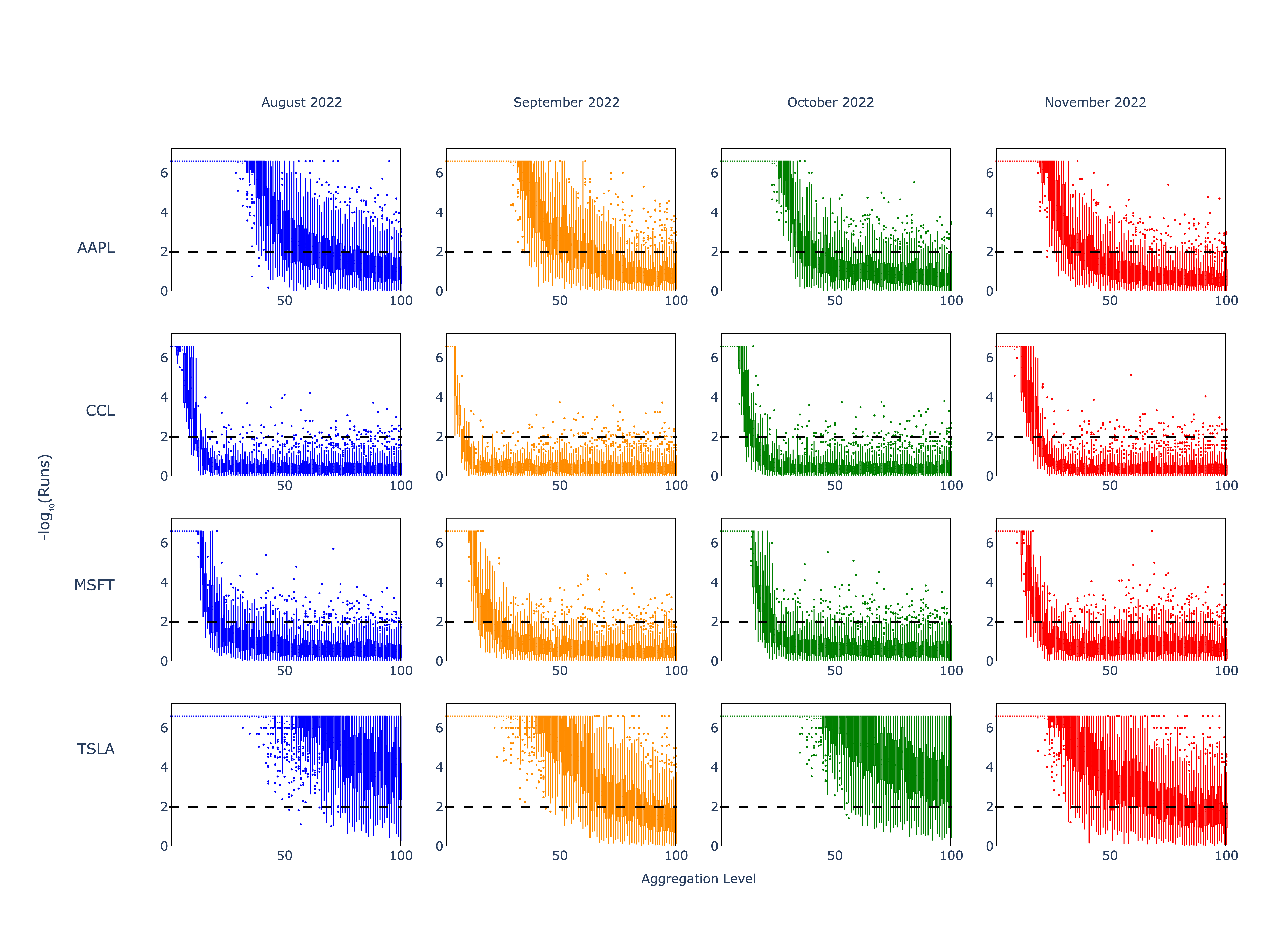}
    \caption{Runs test (NIST STS) applied to AAPL, CCL, MSFT and TSLA: boxplots of $-\log_{10}p$ over the $\ell$ strings at each aggregation level; the dashed line is the threshold $\alpha=0.01$.}
    \label{fig:cases12_runs}
\end{figure}

\begin{figure}[tbp]
    \centering
    \includegraphics[width=0.86\linewidth]{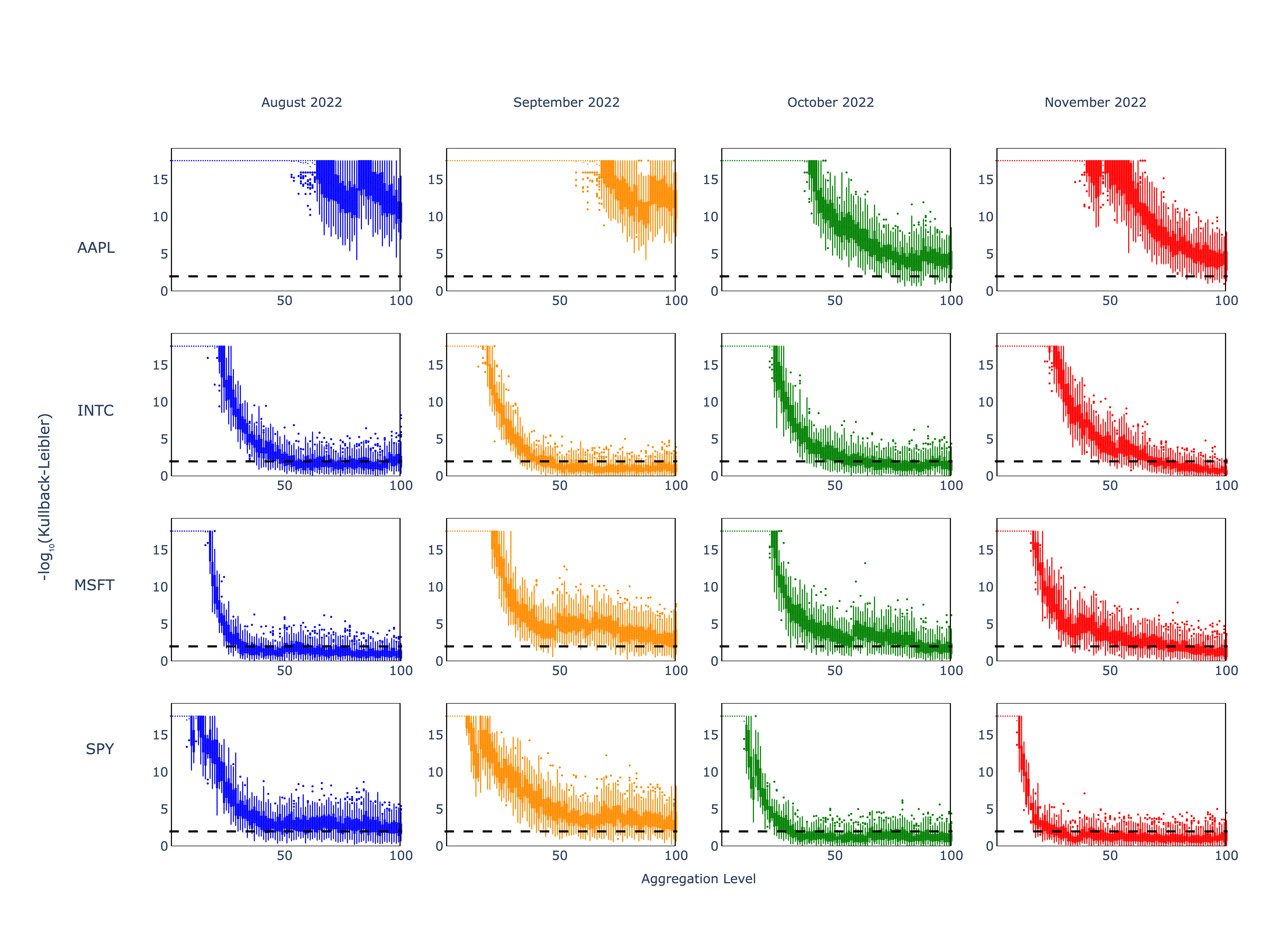}
    \caption{KL test applied to AAPL, INTC, MSFT and SPY (as in Fig.~\ref{fig:cases12_runs}).}
    \label{fig:cases12_kl}
\end{figure}

Figures~\ref{fig:cases12_runs} and \ref{fig:cases12_kl} show the Runs test of NIST STS, which is sensitive to lag-one dependence, and the KL test, which tests independence without requiring balanced frequencies. For all assets and months $-\log_{10}p$ decreases with $\ell$, as expected from the whitening of Section~\ref{subsec:ar1}, and the same holds for most tests of the three batteries (see Supplementary Figs.~S1--S4 for further examples). The level at which strings pass, however, differs markedly between assets. In the Runs test, CCL strings pass from $\ell\approx10$--$20$ and MSFT strings from $\ell\approx15$--$25$, whereas AAPL strings pass only beyond $\ell\approx40$--$90$ and TSLA strings rarely before $\ell=100$. In the KL test, INTC, MSFT and SPY strings become random at $\ell\approx20$--$80$ in most months, while AAPL strings remain non-random up to $\ell=100$ in August and September.

Table~\ref{tab:randomness-threshold} summarises, for each asset, the aggregation level at which strings become and remain random, pooled over tests, batteries and months. The ordering follows the length of the strings: CCL and F, with $\approx5\times10^4$ bits per month, become random at very low levels, whereas AAPL, MSFT, SPY and TSLA, with $0.5$--$1.8\times10^6$ bits per month, need about 100. This is the behaviour predicted by Eq.~\eqref{eq:ell_det}: longer strings make the tests more powerful, so that more aggregation is needed before the residual memory falls below their resolution. The persistent predictability of AAPL and TSLA thus reflects their high trading activity, as anticipated in \cite{shternshis2025price}, but not only that. 
The lag-one autocorrelation of the signs, $\hat q_1(\ell)$, averaged over the $\ell$ strings at each level, makes the comparison quantitative (Supplementary Section~S8, Fig.~S12 and Table~S11). For short-memory models, Eqs.~\eqref{eq:rho_kappa} and \eqref{eq:arcsine} give $\ell q_1(\ell)\to2\kappa/\pi$. For CCL and SNAP, $\ell\hat q_1(\ell)$ levels off at $\approx1.2$--$1.3$ beyond $\ell\approx20$, within its month-to-month scatter, which gives $\kappa\approx2$. With the effective lengths of their records, $N\approx2.4$--$2.9\times10^5$, and $\Lambda^\ast=6.6$, Eq.~\eqref{eq:ell_det} then predicts $\ell_{\rm det}\approx37$--$41$ for the lag-one statistic $n_\ell\hat q_1^{\,2}$, which in the data falls below $\Lambda^\ast$ at $\ell\approx37$--$39$ (median over months). For F and LLY, $\hat q_1$ decays faster and is compatible with zero beyond $\ell\approx30$. For AAPL, INTC, MSFT, SPY and TSLA, by contrast, $\ell\hat q_1(\ell)$ has no plateau: it grows by a factor 1.7--2.8 between $\ell=20$--$40$ and $\ell=60$--$100$, and $\hat q_1(100)=0.03$--$0.06$. The values of $\kappa$ obtained at $\ell=20$--$40$, about 3 for AAPL, INTC and TSLA and 1.3--1.7 for MSFT and SPY, together with $N=0.6$--$3.2\times10^6$, give $\ell_{\rm det}\approx60$--$130$, against $\ell_{\rm det}\leq41$ for the other four assets, which reproduces the ordering of Table~\ref{tab:randomness-threshold}. The slower decay at larger $\ell$ then keeps the lag-one statistic above $\Lambda^\ast$ at $\ell=100$ in 19 of these 20 asset-months. The most active assets thus combine longer records with a memory that is stronger and longer-lived than a single $\kappa$ describes (Section~\ref{subsec:departures}). This also accounts for INTC, which needs high aggregation despite its moderate length: its $\kappa$ is as large as that of AAPL, and in August and September its strings are also unbalanced (Section~\ref{subsubsec:imbalance}). Table~\ref{tab:randomness-threshold} pools tests of very different power and is bounded by $\ell=1$ and $\ell=100$.

\begin{table}[htbp]
  \centering
  \small
  \caption{Aggregation level at which each ticker's bit-strings become and remain random, pooled over all batteries, months and tests. ``Ever random'' is the share of test instances that become and remain random before the maximum level.}
  \label{tab:randomness-threshold}
  \begin{tabular}{lrrrr}
    \toprule
    Ticker & Median & IQR & Mean & Ever random (\%) \\
    \midrule
    F & 1.0 & 1--22 & 24.8 & 91 \\
    CCL & 1.0 & 1--40 & 27.8 & 86 \\
    LLY & 4.5 & 1--94 & 32.1 & 83 \\
    SNAP & 9.0 & 1--91 & 35.6 & 85 \\
    INTC & 99.0 & 1--100 & 59.0 & 53 \\
    SPY & 99.0 & 1--100 & 60.0 & 52 \\
    MSFT & 99.0 & 1--100 & 61.3 & 51 \\
    AAPL & 100.0 & 77--100 & 78.4 & 35 \\
    TSLA & 100.0 & 100--100 & 85.2 & 24 \\
    \bottomrule
  \end{tabular}
\end{table}

Different tests probe different properties, so that a string may look random with respect to one property and not to another. For SPY in August 2022, for instance, the ApproximateEntropy test indicates randomness from $\ell\approx20$, while the MultinomialBitsOverlapping test ($L=4$) still rejects it at $\ell=100$ (Supplementary Figs.~S1--S2).

\subsubsection{Large-tick stocks: maximal predictability at intermediate aggregation}
\label{subsubsec:largetick}

\begin{figure}[tbp]
    \centering
    \includegraphics[width=\linewidth]{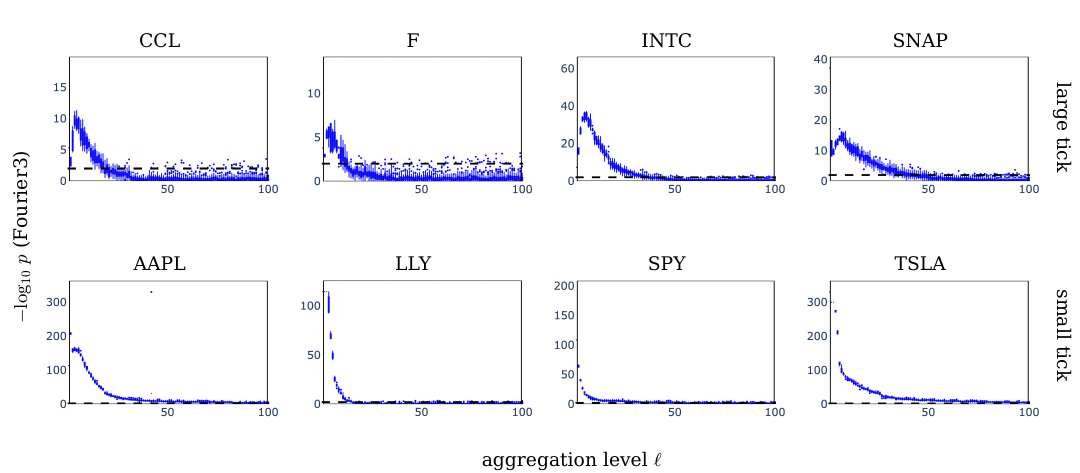}
    \caption{Fourier3 test (Rabbit) in August 2022 for the large-tick stocks CCL, F, INTC and SNAP (top) and for the small-tick assets AAPL, LLY, SPY and TSLA (bottom): boxplots of $-\log_{10}p$ over the $\ell$ strings at each aggregation level; the dashed line is the threshold $\alpha=0.01$. Note the different vertical scales.}
    \label{fig:case3}
\end{figure}

For the four large-tick stocks the Fourier3 spectral test of Rabbit reaches its maximum of $-\log_{10}p$ at $\ell\approx5$--$10$, not at $\ell=1$, before decreasing (Fig.~\ref{fig:case3}, top). For the small-tick assets AAPL, LLY, SPY and TSLA, by contrast, predictability is largest at $\ell=1$ and decreases with aggregation (Fig.~\ref{fig:case3}, bottom); the same contrast holds in the other months (Supplementary Figs.~S3--S4). Similar non-monotonic patterns appear in the AutoCorrelation ($d=1$) test of Rabbit, in the BlockFrequency and CumulativeSums tests of NIST STS and in the entropy-based tests for some assets.

This is the behaviour predicted in Section~\ref{subsec:departures} for a persistent component masked by the bid-ask bounce. Transaction prices of large-tick stocks bounce between the best quotes, which change rarely compared with the rate of transactions, as shown by their large fraction of zero returns (Table~\ref{Table: UHF data}). At the tick level the anti-persistent bounce partly hides the persistence of the price changes; a few transactions of aggregation average the bounce out and expose the persistent component, whose predictability is then removed by further aggregation. The tick-size dependence agrees with the finding of \cite{Calcagnile} that low entropy is associated with a large relative tick size. The mechanism makes a direct prediction: the lag-one autocorrelation of the signs should be negative at $\ell=1$ and positive, with a maximum, at intermediate $\ell$ for large-tick stocks. The data agree (Supplementary Fig.~S12). For CCL, F and SNAP, $\hat q_1(1)$ lies between $-0.08$ and $0.01$, and $\hat q_1(\ell)$ reaches a maximum of $0.06$--$0.12$ at $\ell\approx5$--$13$, the range in which the Fourier3 test is most significant. INTC has a maximum of the same size at $\ell\approx7$--$8$, with $\hat q_1(1)$ close to zero ($-0.02$ to $0.03$). For LLY, MSFT, SPY and TSLA, $\hat q_1(1)$ is positive, between $0.05$ and $0.24$, and $\hat q_1$ is largest at $\ell=1$--$2$; AAPL, which has the largest relative tick size and fraction of zero returns among the small-tick assets, is intermediate, with $\hat q_1(1)=0.03$--$0.09$ and a maximum at $\ell\approx3$--$6$. Merging the executions that share a timestamp, i.e.\ the fills of one aggressive order, into a single transaction lowers the maxima of the large-tick stocks to $0.02$--$0.08$ and moves them to $\ell\approx2$--$7$, while $\hat q_1(1)$ remains negative or close to zero. Periodic algorithmic activity triggered every few transactions could also contribute to these patterns.

\subsubsection{Frequency imbalance}
\label{subsubsec:imbalance}

\begin{figure}[tbp]
  \centering
  \begin{subfigure}[t]{0.32\textwidth}
    \centering
    \includegraphics[width=\linewidth]{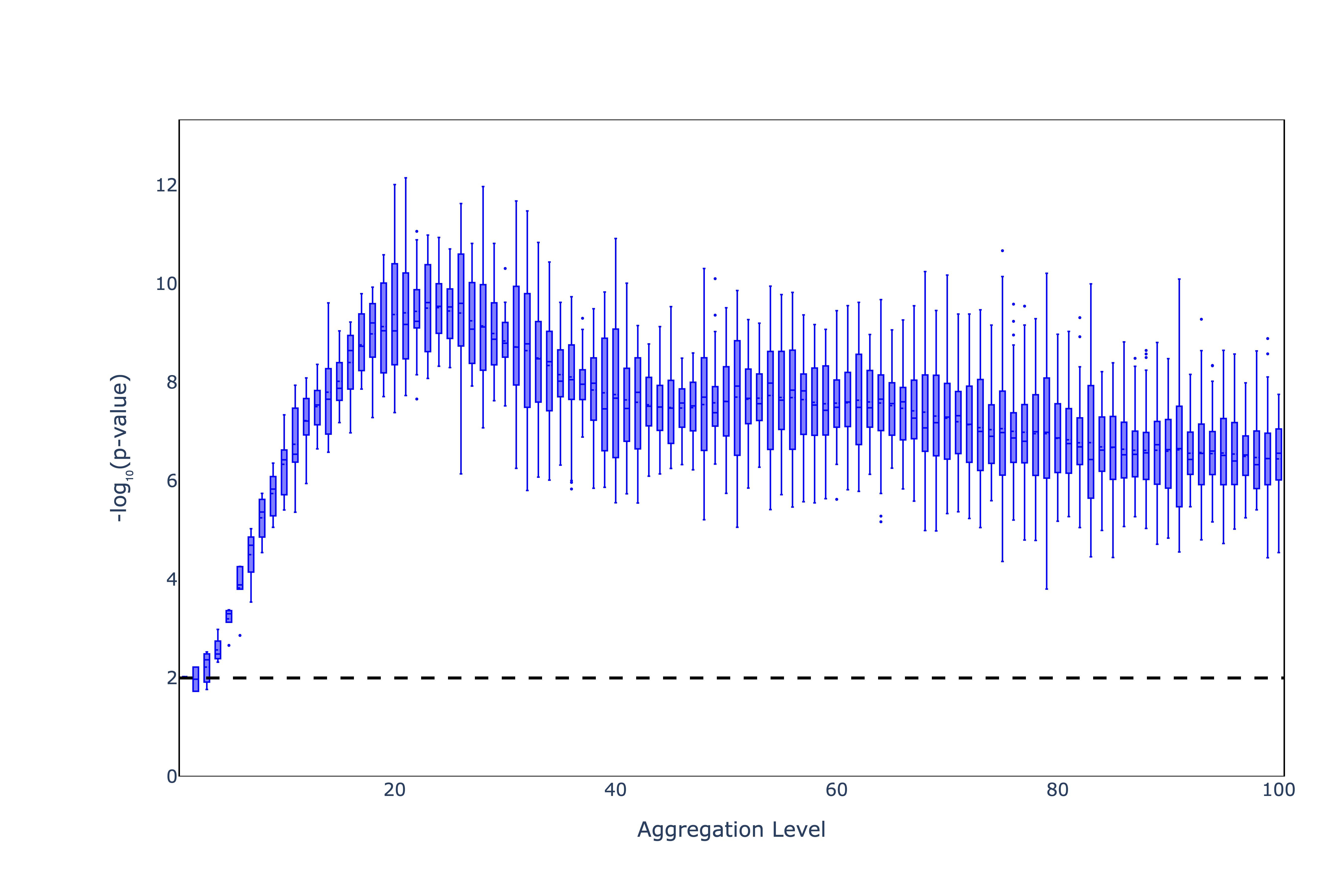}
    \subcaption{Frequency test}
  \end{subfigure}\hfill
  \begin{subfigure}[t]{0.32\textwidth}
    \centering
    \includegraphics[width=\linewidth]{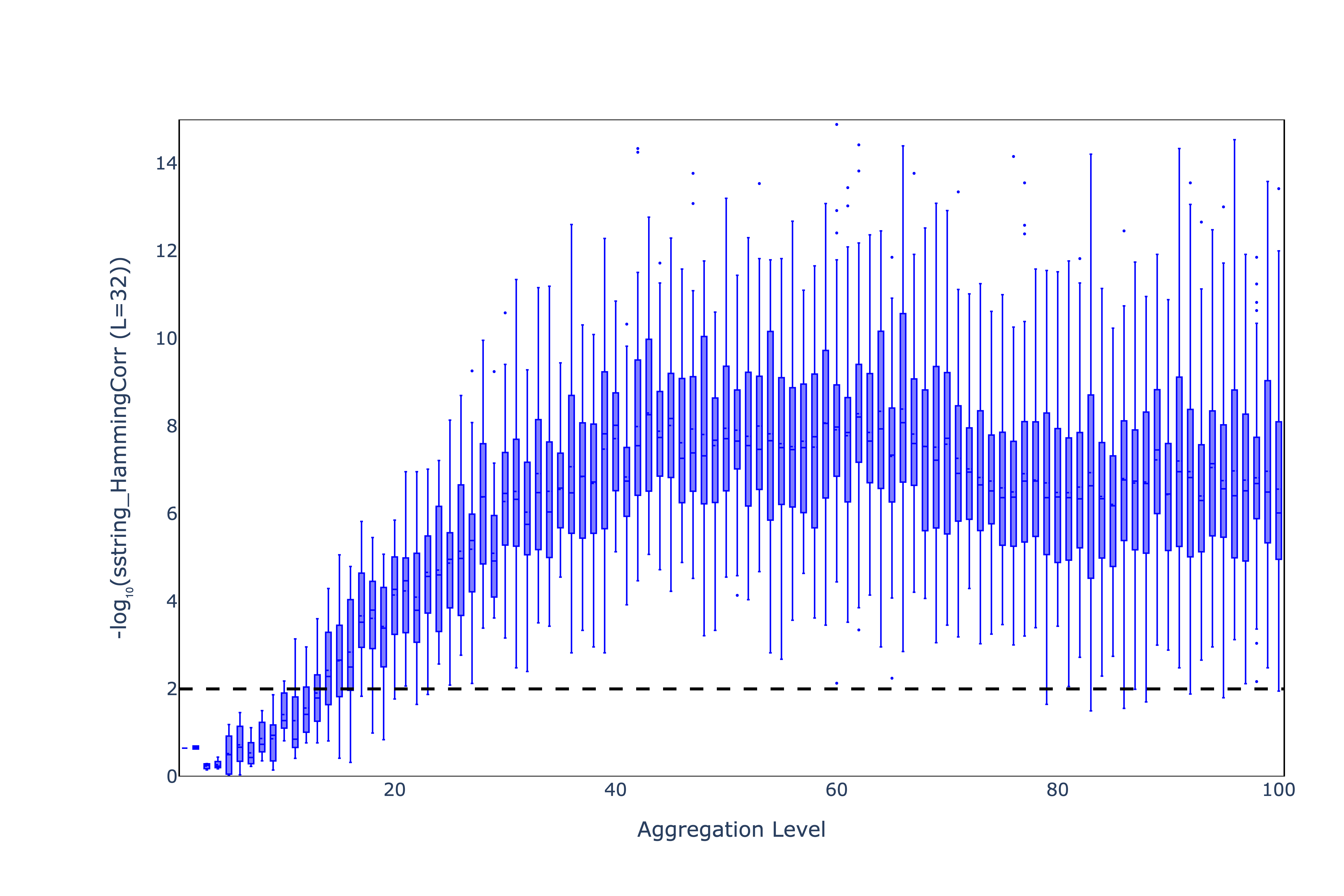}
    \subcaption{HammingCorrelation ($L=32$)}
  \end{subfigure}\hfill
  \begin{subfigure}[t]{0.32\textwidth}
    \centering
    \includegraphics[width=\linewidth]{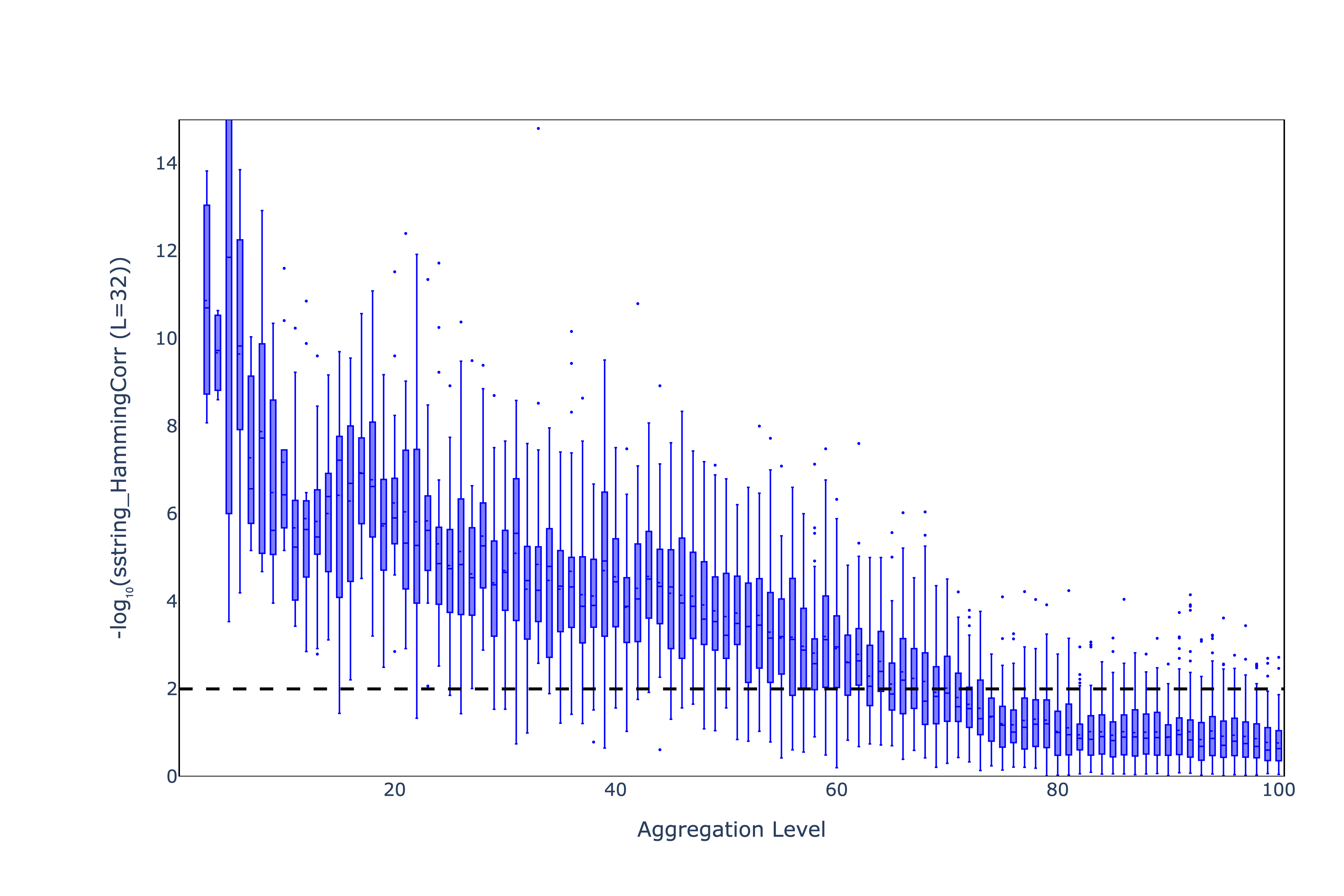}
    \subcaption{HammingCorrelation, balanced}
  \end{subfigure}

  \medskip
  \begin{subfigure}[t]{0.32\textwidth}
    \centering
    \includegraphics[width=\linewidth]{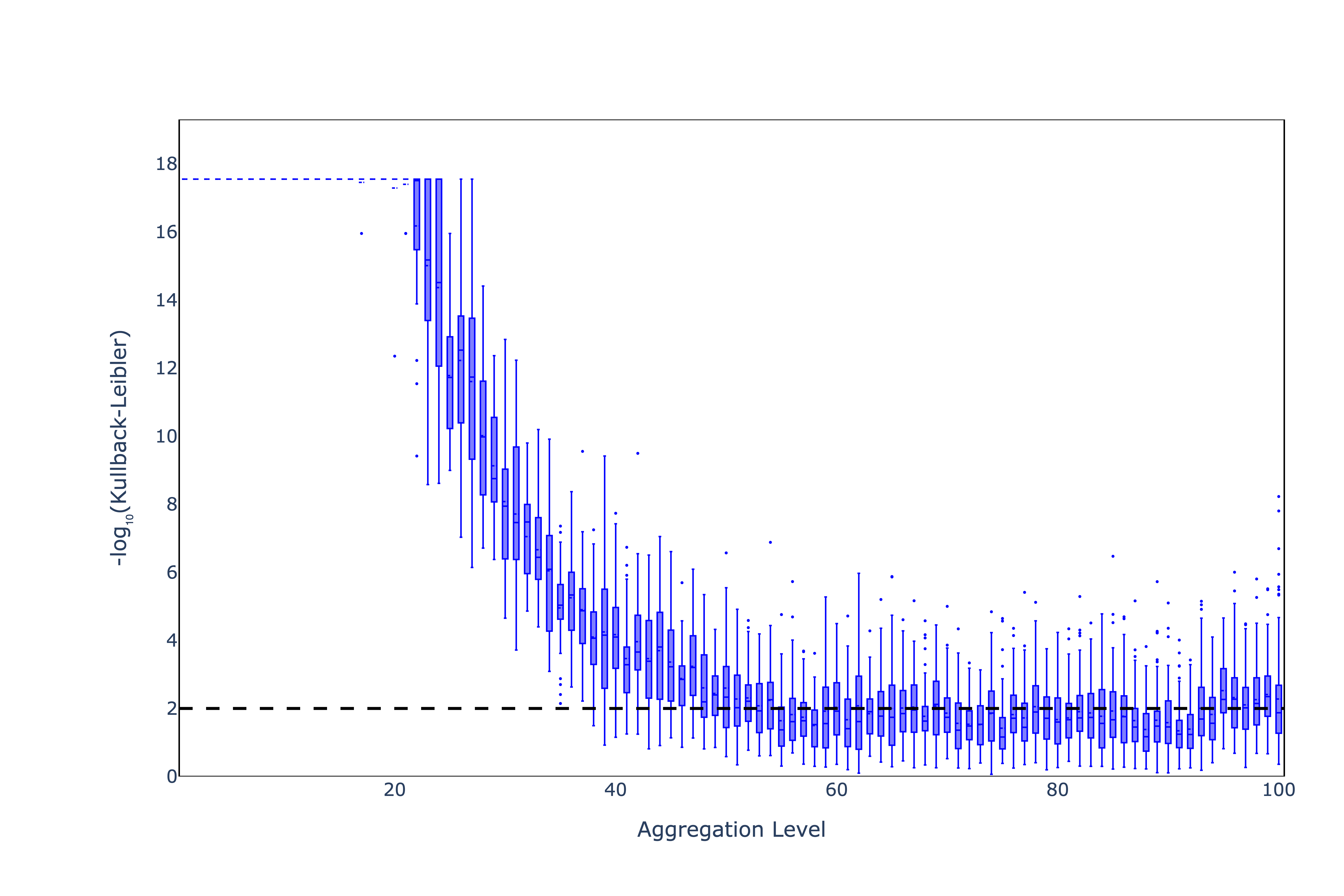}
    \subcaption{KL test}
  \end{subfigure}\qquad
  \begin{subfigure}[t]{0.32\textwidth}
    \centering
    \includegraphics[width=\linewidth]{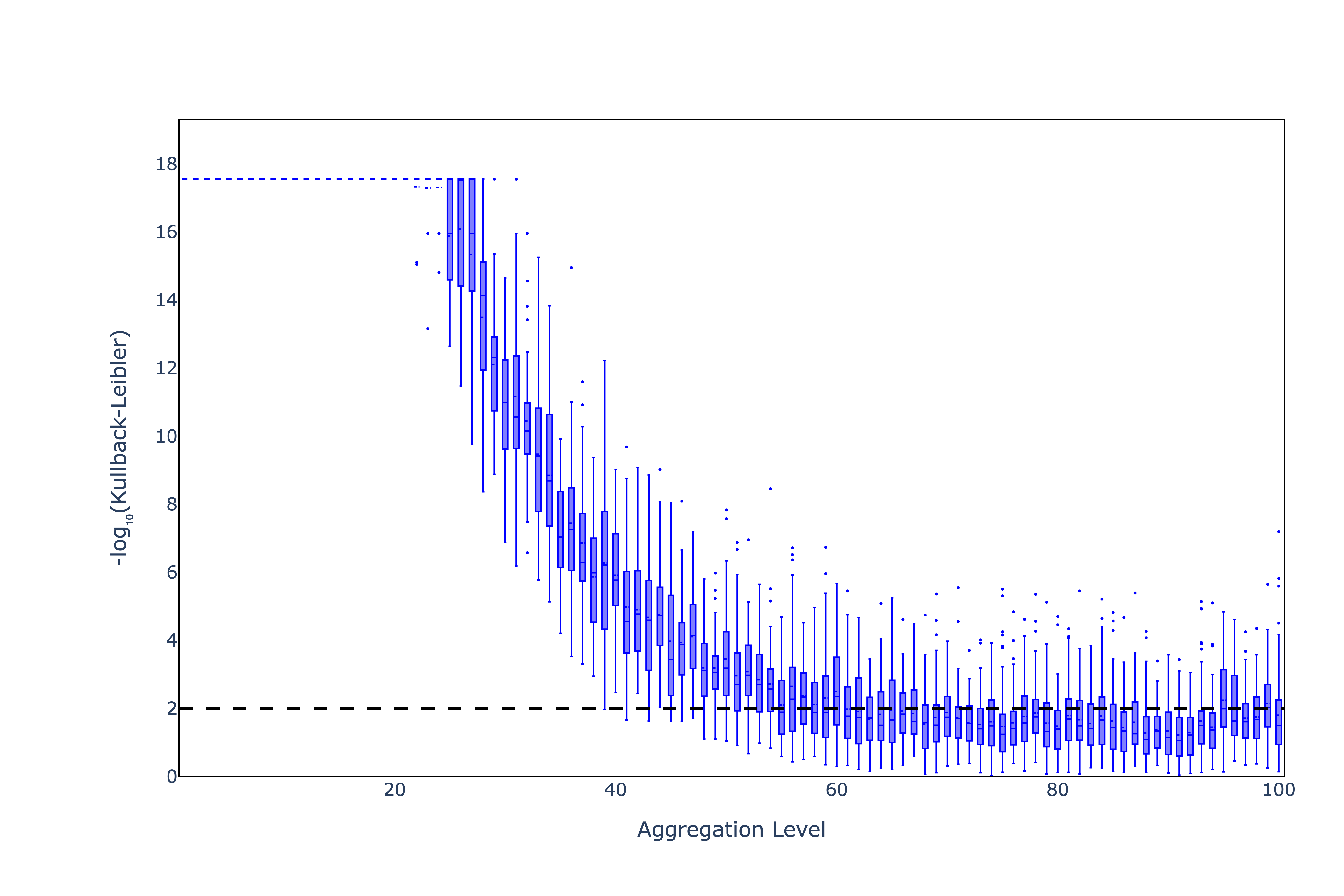}
    \subcaption{KL test, balanced}
  \end{subfigure}
  \caption{INTC, August 2022: (a) frequency test on the statistic $Z$; (b) HammingCorrelation ($L=32$, Rabbit) and (d) KL test on the strings of Eq.~\eqref{Eq:encoding}; (c), (e) the same tests on balanced strings obtained by thresholding at the daily median.}
  \label{fig:case4}
\end{figure}

The HammingCorrelation tests of Alphabit and Rabbit, applied to INTC in August 2022, show predictability \emph{increasing} with aggregation (Fig.~\ref{fig:case4}b). These tests compare the Hamming weights of successive blocks and are therefore sensitive to the balance between zeros and ones. A frequency test, based on the statistic $Z=2|c-N/2|/\sqrt{N}$ for a string of length $N$ with $c$ zeros, shows that the strings are unbalanced and that the imbalance becomes more significant as $\ell$ grows from 1 to about 20, before levelling off (Fig.~\ref{fig:case4}a). In September the imbalance is weaker and significant only for $\ell\gtrsim50$, and in October and November the strings are balanced, with $Z<1.1$ at all levels (Supplementary Fig.~S6, which shows the signed statistic, the monobit $z$-score). Imbalances are not specific to INTC: $Z$ exceeds its critical value, 2.58, in some months also for AAPL, F, LLY, MSFT, SPY and TSLA, and reaches 37 for AAPL and 22 for TSLA, whereas CCL and SNAP strings are balanced in all months (Supplementary Section~S6).

To separate the imbalance from the serial dependence, we build balanced strings by replacing the threshold $r=1$ in Eq.~\eqref{Eq:encoding} with the median $M$ of the daily ratios $P_{j+i\ell}/P_{j+(i-1)\ell}$ (bit 0 if $r<M$, bit 1 if $r>M$). Unlike Eq.~\eqref{Eq:encoding}, this encoding uses the whole trading day and is not online. With balanced strings, the HammingCorrelation test indicates randomness from $\ell\approx70$ (Fig.~\ref{fig:case4}c), while the KL test, which does not require balanced frequencies, indicates randomness from $\ell\approx50$ with either encoding (Fig.~\ref{fig:case4}d,e).

Section~\ref{subsec:departures} identifies two sources of imbalance, a drift and an asymmetry between up and down moves, which differ in size and in their dependence on $\ell$. A drift gives $z\simeq\sqrt{2/\pi}\,T$, where $T=\sum S/(\sum S^2)^{1/2}$ is the realised trend of the string: the sum of its block log-returns, i.e.\ the net intraday log-price change of the month, in units of their realised volatility. In our data $|\sqrt{2/\pi}\,T|<1.6$ for all assets, months and levels, so the trends of the months cannot produce the imbalances observed. For INTC in August, $\sqrt{2/\pi}\,T\approx-0.9$ at all levels, against $z\approx-6$ at $\ell\approx25$ and $-5$ at $\ell=100$. At $\ell=1$ the counts give an exact decomposition. The summed intraday price change is $\Delta=n_+\bar u-n_-\bar d$, with $n_\pm$ the numbers and $\bar u$, $\bar d$ the mean sizes of the up and down moves, so that $n_+-n_-=\Delta/\bar d-n_+(\bar u-\bar d)/\bar d$ is the sum of a trend term and of a size-asymmetry term. For INTC in August, $n_+-n_-=-838$ while $\Delta=-266$ ticks: the down moves were more numerous but smaller on average ($\bar d=0.663$ and $\bar u=0.667$ ticks, most moves being half-tick moves to or from the midpoint, where hidden orders execute), and the two terms are $-401$ and $-437$. For the other assets the asymmetry term dominates: it accounts for at least 70\% of $n_+-n_-$ in all months with $|z|>3$ at $\ell=1$, and for AAPL in October, when the up moves were 3\% larger than the down moves, it gives $z=-11.6$ although $\Delta=+2008$ ticks (Supplementary Table~S10). The imbalances thus reflect mainly the asymmetry between up and down moves, the second mechanism of Section~\ref{subsec:departures}, and they persist when the executions with the same timestamp are merged. As expected for this mechanism, $|z|$ decreases with $\ell$ for most assets beyond $\ell\approx10$, although more slowly than the $1/\ell$ of independent increments. For INTC in August, and for AAPL up to $\ell\approx10$, $|z|$ first grows with $\ell$: as for the drift of Section~\ref{subsec:departures}, the bounce, which dominates the signs at the tick level, masks the asymmetry until aggregation averages the bounce out. Both components of the imbalance are removed by measuring signs relative to the median.

\section{Random numbers from financial time series}
\label{Random numbers from financial time series}

Aggregation turns the non-random strings obtained from tick data into strings that pass the test batteries, for most assets. The procedure is online, since each bit depends only on two past prices, and it requires neither a model nor simulations. Table~\ref{tab:randomness-threshold} shows the aggregation levels at which this happens; the theory of Section~\ref{subsec:finite}, together with the slower decay of the memory of the most active assets (Section~\ref{subsubsec:whitening}), explains why they are higher for more liquid assets.

The amount of randomness that can be extracted is limited. The $\ell$ strings obtained at level $\ell$ are built from the same price path and share substantial mutual information: the entropy per symbol of a sequence obtained by merging several of them in time order falls well below one bit (Supplementary Figs.~S10-S11), so only one of them contributes fresh randomness. A single string at level $\ell$ yields at most about $n_d/\ell$ bits per day, and fewer once zero returns are discarded. For F, with about 13,000 transactions per day and a mean threshold of about 25, this gives a few hundred bits per day. The entropy rates of the strings, estimated with Grassberger's bias correction \cite{Grassberger88} of Eq.~\eqref{naive}, exceed on average $0.98$ bits per symbol for all assets at $\ell=8$ and $0.99$ at $\ell=16$, and approach one bit as $\ell$ grows (Supplementary Table S8). To separate the remaining deficits from the noise of the estimator, we also apply it to random permutations of the strings, which keep the numbers of zeros and ones but have no serial dependence (Supplementary Table S9). The deficits exceed this noise floor by more than two standard deviations of the permutation distribution in all months up to $\ell=8$ for CCL, F, LLY and SNAP, up to $\ell=16$ for INTC, MSFT and SPY, and up to $\ell=100$ for AAPL and TSLA, the same ordering as in Table~\ref{tab:randomness-threshold}. At higher levels they are comparable to the noise of the estimator, whose standard deviation grows from $1.5\times10^{-4}$ bits at $\ell=1$ to $1.8\times10^{-3}$ bits at $\ell=100$.

Passing a battery of tests does not certify randomness: tests can only fail to detect structure, and a cryptographic use would require min-entropy estimates and an analysis of possible manipulation of the source \cite{10.5555/1924892.1924895, Landis25}. Within these limits, aggregated tick data offer a public, verifiable and continuously updated source of pseudorandom bits, which could be classified within the framework of \cite{Machicao21} and used, for instance, in randomness beacons \cite{10.5555/1924892.1924895}.

\section{Conclusions}
\label{Conclusion and future work}

We have treated binarised tick-by-tick prices as the output of a random number generator and studied, with standard test batteries and entropy-based tests, how randomness emerges as transactions are aggregated. We have complemented this with a solvable benchmark: for every stationary short-memory Gaussian model of tick returns, the sign sequence of aggregated returns whitens with an entropy deficit $C_\kappa\ell^{-2}$, controlled by a single number $\kappa$, and in a finite sample its residual memory remains detectable up to an aggregation level $\ell_{\rm det}\propto(\kappa^2N)^{1/3}$.

Theory and experiments agree on several points, most clearly for large-tick stocks.
(i) Randomness increases with aggregation for all assets and in almost all tests, as expected for short-memory microstructural dependence.
(ii) The aggregation level at which strings pass the tests grows with their length: CCL and F become random at much lower levels than AAPL, MSFT, SPY and TSLA. In the benchmark this follows from the growth of the power of the tests with the number of transactions, which turns the qualitative explanation in terms of trading activity into a quantitative one. The lag-one autocorrelation of the signs completes it: where it decays as $1/\ell$, as for CCL and SNAP, the estimated $\kappa\approx2$ predicts the observed detection level, while for the most active assets it decays more slowly, so that their memory, and not only the length of their records, keeps their strings non-random up to $\ell=100$.
(iii) The large-tick stocks CCL, F, INTC and SNAP are most predictable at intermediate aggregation levels, $\ell\approx5$-$10$, close to where the lag-one autocorrelation of their signs, negative or close to zero at $\ell=1$, has its maximum; the small-tick assets are most predictable at $\ell=1$. This is the signature, predicted by the theory, of a persistent component masked by the bid-ask bounce, which dominates the transaction prices of large-tick stocks, and it agrees with the association between low entropy and large relative tick size found in \cite{Calcagnile}.
(iv) Frequency imbalances, strongest for AAPL and TSLA, are removed by measuring signs relative to the median. They are too large to be produced by the price trends of the months, which give $|z|<1.6$, and reflect an asymmetry between the sizes of up and down moves, the second source of imbalance identified by the theory; for most assets their significance decreases with aggregation, as predicted for this mechanism.

Some observations go beyond the short-memory benchmark. For AAPL, INTC, MSFT, SPY and TSLA the lag-one autocorrelation of the signs decays more slowly than $1/\ell$ beyond $\ell\approx20$ and is still $0.03$-$0.06$ at $\ell=100$, where these strings sample the price every 12 to 76~s; AAPL and TSLA strings remain non-random up to $\ell=100$ in the Runs and KL tests. Long-range dependence of returns, for which aggregation does not whiten the signs and the detection level grows as $N$ rather than $N^{1/3}$, and non-stationarity, such as intraday trends, are possible explanations; aggregation over a wider range of levels and the scaling of the detection level with the record length can discriminate between them. Natural extensions are the comparison between transaction and mid-quote prices, which would remove the bounce, and batteries of tests that do not require equiprobable symbols, such as modified frequency tests \citep{Ecuyer99, Rukhin01} and the KL test \cite{shternshis2025price}. Finally, the procedure can be developed into a practical randomness source, such as a randomness beacon based on financial data.

\section*{CRediT authorship contribution statement}

\textbf{Silvia Onofri:} Conceptualization, Data curation, Investigation, Methodology, Software, Validation, Writing -- original draft, Writing -- review \& editing.
\textbf{Andrey Shternshis:} Conceptualization, Data curation, Methodology, Software, Validation, Writing -- original draft, Writing -- review \& editing.
\textbf{Stefano Marmi:} Conceptualization, Methodology, Project administration, Supervision, Writing -- review \& editing.

\section*{Declaration of competing interest}

The authors declare that they have no known competing financial interests or personal relationships that could have appeared to influence the work reported in this paper.

\section*{Data availability}

The tick data were obtained from LOBSTER (\url{www.lobsterdata.com}) under a licence that does not allow redistribution. The code used for the analysis and for the figures is available at \url{https://github.com/silviaonofri/Emergence-of-Randomness-in-Temporally-Aggregated-Financial-Tick-Sequences}, together with the Supplementary Material.

\section*{Declaration of generative AI and AI-assisted technologies in the manuscript preparation process}

During the preparation of this work the authors used Claude (Anthropic) in order to assist with the derivation and numerical verification of the results of Section~\ref{Theoretical Foundations}, the production of Fig.~\ref{fig:theory} and the composition of Fig.~\ref{fig:case3}, and language editing. After using this tool, the authors reviewed and edited the content as needed and take full responsibility for the content of the published article.

\section*{Acknowledgements}
We thank Prof. Fabrizio Lillo for the helpful discussions and suggestions. We are also grateful to the participants of Financial Cryptography and Data Security 2026 (FC 2026) and DeFi \& Crypto 2026 for their valuable comments on an earlier version of this paper.

\bibliographystyle{elsarticle-num}
\bibliography{bib}

\end{document}